# MINDDESKTOP: A GENERAL PURPOSE BRAIN COMPUTER INTERFACE

Ori Ossmy, Ofir Tam, Rami Puzis, Lior Rokach, Ohad Inbar, Yuval Elovici

*Telekom Innovation Laboratories and*
*Software & Information Systems Engineering Dept., Ben-Gurion University*
*Beer Sheva, 84105, Israel*
oriossmy@gmail.com, tof@bgu.ac.il, puzis@bgu.ac.il, liorrk@bgu.ac.il, inbaro@bgu.ac.il, elovici@bgu.ac.il

Abstract: Recent advances in electroencephalography (EEG) and electromyography (EMG) enable communication for people with severe disabilities. In this paper we present a system that enables the use of regular computers using an off-the-shelf EEG/EMG headset, providing a pointing device and virtual keyboard that can be used to operate any Windows based system, minimizing the user effort required for interacting with a personal computer. Effectiveness of the proposed system is evaluated by a usability study, indicating decreasing learning curve for completing various tasks. The proposed system is available in the link provided.

## 1 INTRODUCTION

We are all familiar with the ubiquity of the personal computer (PC) in modern society. Unfortunately, disabilities prevent many people from standard interaction with a PC. Many research activities have been undertaken in recent years to develop technologies to enable efficient brain computer interfaces (BCI) through the analysis of ongoing electro-encephalographic (EEG) signal of the user (Felzer et al., 2002; Wolpaw et al., 2002), EEG based systems require no physical activity, thus enabling even users suffering from complete paralysis but normal cognition (also known as locked-in syndrome, Smith and Delargy, 2005) to communicate using BCI (Markand, 1976).

Research and commercial applications of EEG based BCI gained significant traction after the introduction of commercial, off-the-shelf devices for measuring muscular and brain activity, like the Emotiv EPOC and Insight headsets[1] and their software suites that can produce a signal each time a predefined cognitive or facial activity is detected.

However, two problems are common to EEG based BCI systems: speed and sensitivity to noise. Neuper et al. (2003) for example, refer to a typing rate of just one character per minute. As we report in this paper, MindDesktop[2,3] achieved a typing rate of approximately 20 seconds per character. Mugler et al. (2014) report a transfer rate of 3 bits per second using ECoG implants. This roughly corresponds to 36 characters per minute in the 26 letters English alphabet with additional six punctuation marks or special operations.

Sensitivity to noise (and thus errors) results from the amplification of the tiny potentials comprising the EEG (Goncharova et al., 2003). EMG can be used to handle EEG noise (Doherty et al., 2000; Barreto et al., 1999), yet untrained users may find it difficult to use these systems quickly and accurately.

Neural readings are significantly improved by the use of nano implants (Seo et al., 2016). Several companies such as Neuralink, Kernel, and Facebook invest in BCI technologies with the objective to enhance the human-machine interaction in the everyday use. Facebook researchers expect to demonstrate a BCI that makes it possible to deliver hundred words per minute (five times faster that typing on a smartphone)[3].

In this paper we propose MindDesktop, an enabling system based on adaptive EEG/EMG devices, whose primary goal is to improve computer accessibility for people with severe disabilities. The contribution of this paper is twofold: First, we developed a generic enabling layer to bridge between a Windows-operated PC and input devices. Although our evaluation was done with the Emotiv EPOC headset, any input device may be used as long as it has the ability to communicate three distinct signals in a user-controlled manner. These signals are interpreted by MindDesktop as mouse clicks or keystrokes according to the system context. The second contribution is the hierarchical UI design approach we suggest for implementing a virtual keyboard and virtual pointing device. This design minimizes both errors and the number of actions required to complete a task. Use of the same approach for both input methods enables a user to operate the computer with a single accessibility device.

---

1. https://www.emotiv.com/
2. https://www.youtube.com/watch?v=bawYa4yrgSE
3. http://www.ise.bgu.ac.il/faculty/liorr/MindDesktop1.zip (password: bgulab).
4. https://www.engadget.com/2017/04/19/facebook-details-its-plans-for-a-brain-computer-interface/

A short usability study evaluated the system's effectiveness by measuring the number of actions a user needs to perform in order to accomplish a specific a set of tasks.

## 2 BACKGROUND ON THE EMOTIV EPOC HEADSET

The Emotiv EEG headset is equipped with 14 saline sensors to sample brain activity. The headset connects with a Windows operated PC via WiFi, thus allowing the computer to be arbitrarily positioned next to the user. The headset includes three major software suites: cognitive, expressive and affective. The cognitive suite can be trained to detect specific cognitive actions by recording the brain activity during a training process. Later, it is able to detect this activity and generate its corresponding signal. Each signal is received and processed by MindDesktop, which translates it to mouse clicks, keystrokes, or application specific operations (e.g. play or pause). Operation of the cognitive suite requires significant skill and effort, especially as the number of cognitive actions increases. We have limited the number of actions required to operate MindDesktop to three, in order to reduce the error rate during detection, yet maintain a functional interface.

The second, expressive suite, can detect the user's facial movements, e.g., moving one's teeth or raising eyebrows. We refer to these expressions as expressive actions. In contrast to cognitive actions, the interpretation of which depends entirely on the user, expressive actions are predefined. Overall, the accuracy of detecting expressive action is higher than that of detecting cognitive actions and the effort required to invoke expressive action is lower. A training process can further increase detection accuracy.

## 3. MINDDESKTOP ARCHITECTURE

MindDesktop consists of three major components as depicted in Figure 1: Core, UI, and Database.

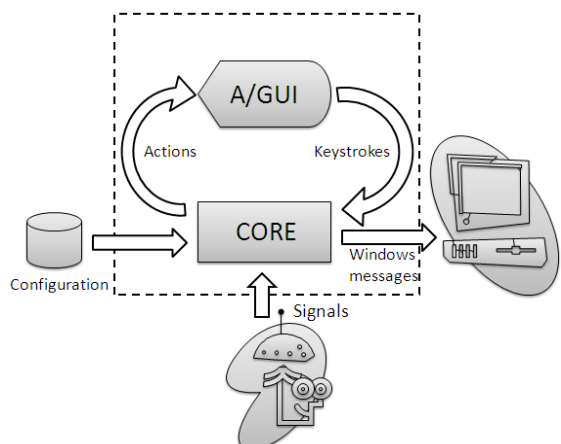

Figure 1 – MindDesktop high level architecture.

**Core component:** This component is responsible for interaction between the input device, the UI, the operating system, and the internal storage and configuration. Each new signal that passes the detection threshold is interpreted as a user action and is communicated to the UI component, which communicates a keystroke or mouse click back to the core, where it is dispatched to the current application.

**UI component:** The UI component manages the system's graphical interface by visualizing the state of virtual input devices (keyboard or pointing device) and providing visual or auditory feedback. Once the actual input is selected (e.g. a letter is chosen on a virtual keyboard) it is communicated back to the core.

**Database component:** The database component contains a full description of the user's profile and system configuration. During the configuration process, the user can choose the layout of an input device for each application and customize shortcut keys. The database also contains a predictive text dictionary and a set of sounds.

## 4. VIRTUAL INPUT DEVICES

MindDesktop includes two virtual input devices - a pointing device and a keyboard. Both utilize the same UI scheme and maintain a tree-like hierarchy.

Both are controlled by just three user actions, yet strive to minimize the number of actions required to complete an interaction with a PC. The root of the tree is the entry point to each input device. Users navigate through the hierarchy using three actions: scroll, zoom in, and zoom out as depicted in Figure 2. Scrolling changes the virtual device states on the current hierarchy level (i.e. select a sibling). Zooming in expands the current state allowing the user to scroll through the children. The actual keystrokes and clicks are on the lowest level of the hierarchy (leaves). Zooming in on a leaf communicates the selected input to the operating system. Zooming out collapses the current state and expands the parent of the hierarchy. In the root state, zooming out cancels the operation.

scroll between them and choose the one that covers the spot of interest. Zooming-in splits the selected screen area into four smaller areas. This process is repeated a preconfigured number of times, depending on the desired selection accuracy. The lowest hierarchy level is indicated by a target and a special sound. Once the spot of interest is reached, the user can either click or double-click by performing one or two additional zoom-in actions. After the first zoom-in, the target rectangle blinks for 4 seconds. Zooming in again during these 4 seconds executes a double-click; otherwise, a single click is communicated to the operating system. Zooming out reverses all navigation and, at the root level, closes the pointing device and activates the virtual keyboard.

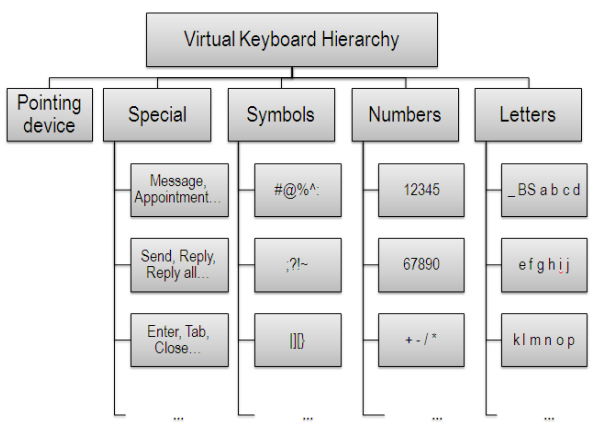

Figure 2: Input device state hierarchy.

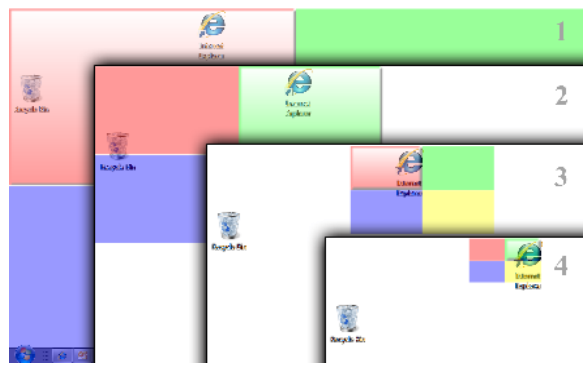

Figure 3: Pointing device UI.

## 4.1 Hierarchical Keyboard

The virtual keyboard has three levels: The root provides access to groups of keys (shortcuts, symbols, number, letters, and desktop); the second displays up to five subgroups; and the third displays the keys themselves (up to six keys per group). Navigation between the groups and the selection of a specific key is achieved by zooming in.

## 4.2 Hierarchical Pointing Device

In addition to a virtual keyboard, a pointing device is necessary to enable users to choose objects on the screen (e.g., an icon or button). We chose to design the interaction of the pointing device in the same way as that of the keyboard. The suggested pointing device is visualized as four, partially transparent, different colored rectangles laid over the screen (see Figure 3). The rectangles divide the screen into four equal parts, allowing the user to

# 5. USABILITY EVALUATION

## 5.1 Participants and apparatus

A usability evaluation was performed in order to evaluate MindDesktop. The study included 17 healthy PC users aged 21 to 55 (avg. 30.6), 8 males and 9 females, and was performed on a standard PC laptop – Lenovo T400.

Each user underwent a short tutorial on the use of MindDesktop, and the Emotiv expressive suite was trained to detect his/her facial expressions – using left and right smirk and a smile. Next, the users participated in three testing sessions (one per day). In each session they were asked to perform five different tasks. 1) Use the pointing device to click on a spot that appears in a random location on the screen. 2) Use the pointing device to open a folder. 3) Open Windows Media Player and play a video located on the Media Player bar; stop, rewind, or pause the video using a Media Player keyboard layout. 4) Open Internet Explorer and search for a short keyword using a search engine. 5) Open Microsoft Outlook and send a new email with a one word subject and 12 characters content. Finally, close Outlook using the pointing device. Tasks were slightly modified in each session in order to avoid over-training.

During each session the time required to complete a task and the number of user actions performed were measured. The latter was compared to the minimal number of user actions required to complete the task. After each session, users completed three System Usability Scale (SUS) questionnaires (Brooke, 1996) to evaluate their satisfaction from the virtual keyboard, the pointing device, and the system in general.

## 5.2 Results and discussions

Figure 4 presents the mean time required to complete the tasks. Vertical bars denote 0.95 confidence intervals. Reduced time required to perform tasks in subsequent sessions, demonstrates a learning curve. To examine the effects of the session number on the time required to complete the task, a two-way ANOVA with repeated measures was performed. The dependent variable was the mean time. The results show a main effect of the number of sessions $F(2,219)=9.72$, $p<0.001$. In addition, males are about 25% faster than females, $F(1,219)=11.38$, $p<0.001$. We assume that hair length is a contributing factor for the reported difference, as males' hair is shorter and therefore the EPOC headset works better on them. The interaction effect of gender and session was also statistically significant with $F(2,219)=3.38$, $p<0.05$, showing an improvement among the females to be far more noticeable. By the third session, the difference between genders is almost eliminated, meaning that if long hair was the problem, it was overcome quickly.

The task variable is also statistically significant with $F(4,219)=111.32$, $p<0.001$. Specifically tasks 4 and 5 were considered to be much more complicated. In fact, 4 out of 17 participants did not succeed in completing task 5 in the first session. The results also indicate that there is a clear positive correlation between the time spent, the total number of actions, and the number of excess actions ($r^2$ is 0.88 and 0.75 respectively). Age had no effect on performance.

Figure 5 presents the SUS score as reported by users for the keyboard, pointing device and the overall system. In all three cases, the SUS significantly increases with the number of sessions, $F(2,144)=11.41$, $p<.001$. Moreover, user satisfaction from the system and the pointing device is

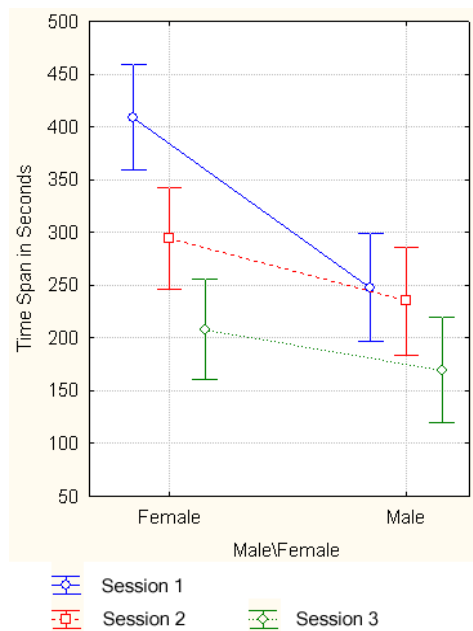

Figure 4: The time span of males and females in three sessions.

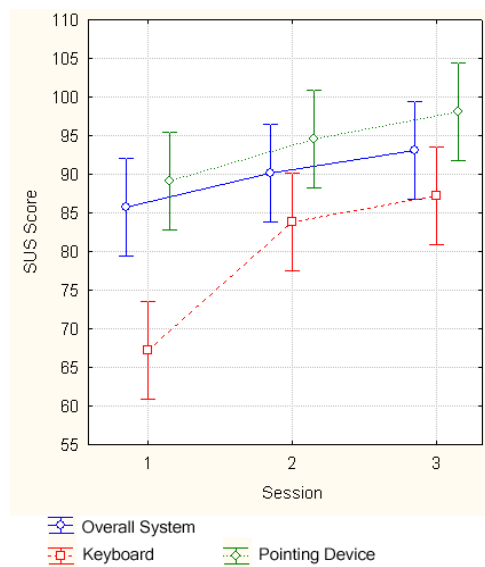

Figure 5: The SUS score reported by the users.

significantly higher than the satisfaction from the keyboard ($F(2,144)=16.2$, $p<0.001$). It is encouraging that the differences between the various components decrease with the number of sessions. While users are initially not satisfied with the keyboard, with short practice their attitude significantly improves, as well as their objective performance metrics. In the third session, all users successfully completed the fifth task (sending an

email) in less than 13 minutes. Four users managed to complete this task in about 4 minutes.

## 6. CONCLUSIONS AND FUTURE WORK

In this paper we studied whether an EEG-based HCI can be used to operate PCs. For this purpose we developed a new hierarchical pointing device and virtual keyboard and examined their performance in a usability study. The results indicate that users can quickly learn how to activate the new interface and efficiently use it to operate a PC.

We believe that hierarchical interfaces and the pointing device presented in this study are beneficial for variety of input technologies including eye tracking and future BCI.